\documentclass[twocolumn,letterpaper,aps,prl,longbibliography,superscriptaddress,showpacs,floatfix]{revtex4-1}

\usepackage{graphicx}	
\usepackage{xspace}	

\begin{document}

\title{ Inclusive cross section and double-helicity asymmetry 
for $\pi^{0}$ production at midrapidity in $p$$+$$p$ collisions at 
$\sqrt{s}=510$ GeV }

\newcommand{\abilene}{Abilene Christian University, Abilene, Texas 79699, USA}
\newcommand{\augie}{Department of Physics, Augustana University, Sioux Falls, South Dakota 57197, USA}
\newcommand{\banaras}{Department of Physics, Banaras Hindu University, Varanasi 221005, India}
\newcommand{\barc}{Bhabha Atomic Research Centre, Bombay 400 085, India}
\newcommand{\baruch}{Baruch College, City University of New York, New York, New York, 10010 USA}
\newcommand{\bnlcoll}{Collider-Accelerator Department, Brookhaven National Laboratory, Upton, New York 11973-5000, USA}
\newcommand{\bnlphys}{Physics Department, Brookhaven National Laboratory, Upton, New York 11973-5000, USA}
\newcommand{\caucr}{University of California-Riverside, Riverside, California 92521, USA}
\newcommand{\charlesczech}{Charles University, Ovocn\'{y} trh 5, Praha 1, 116 36, Prague, Czech Republic}
\newcommand{\chonbuk}{Chonbuk National University, Jeonju, 561-756, Korea}
\newcommand{\ciae}{Science and Technology on Nuclear Data Laboratory, China Institute of Atomic Energy, Beijing 102413, P.~R.~China}
\newcommand{\cns}{Center for Nuclear Study, Graduate School of Science, University of Tokyo, 7-3-1 Hongo, Bunkyo, Tokyo 113-0033, Japan}
\newcommand{\colorado}{University of Colorado, Boulder, Colorado 80309, USA}
\newcommand{\columbia}{Columbia University, New York, New York 10027 and Nevis Laboratories, Irvington, New York 10533, USA}
\newcommand{\czechtech}{Czech Technical University, Zikova 4, 166 36 Prague 6, Czech Republic}
\newcommand{\elte}{ELTE, E{\"o}tv{\"o}s Lor{\'a}nd University, H-1117 Budapest, P{\'a}zm{\'a}ny P.~s.~1/A, Hungary}
\newcommand{\ewha}{Ewha Womans University, Seoul 120-750, Korea}
\newcommand{\fsu}{Florida State University, Tallahassee, Florida 32306, USA}
\newcommand{\gsu}{Georgia State University, Atlanta, Georgia 30303, USA}
\newcommand{\hanyang}{Hanyang University, Seoul 133-792, Korea}
\newcommand{\hiroshima}{Hiroshima University, Kagamiyama, Higashi-Hiroshima 739-8526, Japan}
\newcommand{\howard}{Department of Physics and Astronomy, Howard University, Washington, DC 20059, USA}
\newcommand{\ihepprot}{IHEP Protvino, State Research Center of Russian Federation, Institute for High Energy Physics, Protvino, 142281, Russia}
\newcommand{\illuiuc}{University of Illinois at Urbana-Champaign, Urbana, Illinois 61801, USA}
\newcommand{\inrras}{Institute for Nuclear Research of the Russian Academy of Sciences, prospekt 60-letiya Oktyabrya 7a, Moscow 117312, Russia}
\newcommand{\instpasczech}{Institute of Physics, Academy of Sciences of the Czech Republic, Na Slovance 2, 182 21 Prague 8, Czech Republic}
\newcommand{\isu}{Iowa State University, Ames, Iowa 50011, USA}
\newcommand{\jaea}{Advanced Science Research Center, Japan Atomic Energy Agency, 2-4 Shirakata Shirane, Tokai-mura, Naka-gun, Ibaraki-ken 319-1195, Japan}
\newcommand{\jyvaskyla}{Helsinki Institute of Physics and University of Jyv{\"a}skyl{\"a}, P.O.Box 35, FI-40014 Jyv{\"a}skyl{\"a}, Finland}
\newcommand{\karoly}{K\'aroly R\'oberts University College, H-3200 Gy\"ngy\"os, M\'atrai\'ut 36, Hungary}
\newcommand{\kek}{KEK, High Energy Accelerator Research Organization, Tsukuba, Ibaraki 305-0801, Japan}
\newcommand{\korea}{Korea University, Seoul, 136-701, Korea}
\newcommand{\kurchatov}{National Research Center ``Kurchatov Institute", Moscow, 123098 Russia}
\newcommand{\kyoto}{Kyoto University, Kyoto 606-8502, Japan}
\newcommand{\labllr}{Laboratoire Leprince-Ringuet, Ecole Polytechnique, CNRS-IN2P3, Route de Saclay, F-91128, Palaiseau, France}
\newcommand{\lahorelums}{Physics Department, Lahore University of Management Sciences, Lahore 54792, Pakistan}
\newcommand{\lawllnl}{Lawrence Livermore National Laboratory, Livermore, California 94550, USA}
\newcommand{\losalamos}{Los Alamos National Laboratory, Los Alamos, New Mexico 87545, USA}
\newcommand{\lund}{Department of Physics, Lund University, Box 118, SE-221 00 Lund, Sweden}
\newcommand{\maryland}{University of Maryland, College Park, Maryland 20742, USA}
\newcommand{\mass}{Department of Physics, University of Massachusetts, Amherst, Massachusetts 01003-9337, USA}
\newcommand{\michigan}{Department of Physics, University of Michigan, Ann Arbor, Michigan 48109-1040, USA}
\newcommand{\muhlenberg}{Muhlenberg College, Allentown, Pennsylvania 18104-5586, USA}
\newcommand{\myongji}{Myongji University, Yongin, Kyonggido 449-728, Korea}
\newcommand{\nagasaki}{Nagasaki Institute of Applied Science, Nagasaki-shi, Nagasaki 851-0193, Japan}
\newcommand{\nara}{Nara Women's University, Kita-uoya Nishi-machi Nara 630-8506, Japan}
\newcommand{\natmephi}{National Research Nuclear University, MEPhI, Moscow Engineering Physics Institute, Moscow, 115409, Russia}
\newcommand{\newmex}{University of New Mexico, Albuquerque, New Mexico 87131, USA}
\newcommand{\nmsu}{New Mexico State University, Las Cruces, New Mexico 88003, USA}
\newcommand{\ohio}{Department of Physics and Astronomy, Ohio University, Athens, Ohio 45701, USA}
\newcommand{\ornl}{Oak Ridge National Laboratory, Oak Ridge, Tennessee 37831, USA}
\newcommand{\orsay}{IPN-Orsay, Univ. Paris-Sud, CNRS/IN2P3, Universit\'e Paris-Saclay, BP1, F-91406, Orsay, France}
\newcommand{\peking}{Peking University, Beijing 100871, P.~R.~China}
\newcommand{\pnpi}{PNPI, Petersburg Nuclear Physics Institute, Gatchina, Leningrad region, 188300, Russia}
\newcommand{\riken}{RIKEN Nishina Center for Accelerator-Based Science, Wako, Saitama 351-0198, Japan}
\newcommand{\rikjrbrc}{RIKEN BNL Research Center, Brookhaven National Laboratory, Upton, New York 11973-5000, USA}
\newcommand{\rikkyo}{Physics Department, Rikkyo University, 3-34-1 Nishi-Ikebukuro, Toshima, Tokyo 171-8501, Japan}
\newcommand{\saispbstu}{Saint Petersburg State Polytechnic University, St.~Petersburg, 195251 Russia}
\newcommand{\seoulnat}{Department of Physics and Astronomy, Seoul National University, Seoul 151-742, Korea}
\newcommand{\stonybrkc}{Chemistry Department, Stony Brook University, SUNY, Stony Brook, New York 11794-3400, USA}
\newcommand{\stonycrkp}{Department of Physics and Astronomy, Stony Brook University, SUNY, Stony Brook, New York 11794-3800, USA}
\newcommand{\tenn}{University of Tennessee, Knoxville, Tennessee 37996, USA}
\newcommand{\titech}{Department of Physics, Tokyo Institute of Technology, Oh-okayama, Meguro, Tokyo 152-8551, Japan}
\newcommand{\tsukuba}{Center for Integrated Research in Fundamental Science and Engineering, University of Tsukuba, Tsukuba, Ibaraki 305, Japan}
\newcommand{\vandy}{Vanderbilt University, Nashville, Tennessee 37235, USA}
\newcommand{\weizmann}{Weizmann Institute, Rehovot 76100, Israel}
\newcommand{\wigner}{Institute for Particle and Nuclear Physics, Wigner Research Centre for Physics, Hungarian Academy of Sciences (Wigner RCP, RMKI) H-1525 Budapest 114, POBox 49, Budapest, Hungary}
\newcommand{\yonsei}{Yonsei University, IPAP, Seoul 120-749, Korea}
\newcommand{\zagreb}{University of Zagreb, Faculty of Science, Department of Physics, Bijeni\v{c}ka 32, HR-10002 Zagreb, Croatia}
\affiliation{\abilene}
\affiliation{\augie}
\affiliation{\banaras}
\affiliation{\barc}
\affiliation{\baruch}
\affiliation{\bnlcoll}
\affiliation{\bnlphys}
\affiliation{\caucr}
\affiliation{\charlesczech}
\affiliation{\chonbuk}
\affiliation{\ciae}
\affiliation{\cns}
\affiliation{\colorado}
\affiliation{\columbia}
\affiliation{\czechtech}
\affiliation{\elte}
\affiliation{\ewha}
\affiliation{\fsu}
\affiliation{\gsu}
\affiliation{\hanyang}
\affiliation{\hiroshima}
\affiliation{\howard}
\affiliation{\ihepprot}
\affiliation{\illuiuc}
\affiliation{\inrras}
\affiliation{\instpasczech}
\affiliation{\isu}
\affiliation{\jaea}
\affiliation{\jyvaskyla}
\affiliation{\karoly}
\affiliation{\kek}
\affiliation{\korea}
\affiliation{\kurchatov}
\affiliation{\kyoto}
\affiliation{\labllr}
\affiliation{\lahorelums}
\affiliation{\lawllnl}
\affiliation{\losalamos}
\affiliation{\lund}
\affiliation{\maryland}
\affiliation{\mass}
\affiliation{\michigan}
\affiliation{\muhlenberg}
\affiliation{\myongji}
\affiliation{\nagasaki}
\affiliation{\nara}
\affiliation{\natmephi}
\affiliation{\newmex}
\affiliation{\nmsu}
\affiliation{\ohio}
\affiliation{\ornl}
\affiliation{\orsay}
\affiliation{\peking}
\affiliation{\pnpi}
\affiliation{\riken}
\affiliation{\rikjrbrc}
\affiliation{\rikkyo}
\affiliation{\saispbstu}
\affiliation{\seoulnat}
\affiliation{\stonybrkc}
\affiliation{\stonycrkp}
\affiliation{\tenn}
\affiliation{\titech}
\affiliation{\tsukuba}
\affiliation{\vandy}
\affiliation{\weizmann}
\affiliation{\wigner}
\affiliation{\yonsei}
\affiliation{\zagreb}
\author{A.~Adare} \affiliation{\colorado} 
\author{C.~Aidala} \affiliation{\losalamos} \affiliation{\michigan} 
\author{N.N.~Ajitanand} \affiliation{\stonybrkc} 
\author{Y.~Akiba} \affiliation{\riken} \affiliation{\rikjrbrc} 
\author{R.~Akimoto} \affiliation{\cns} 
\author{J.~Alexander} \affiliation{\stonybrkc} 
\author{M.~Alfred} \affiliation{\howard} 
\author{K.~Aoki} \affiliation{\kek} \affiliation{\riken} 
\author{N.~Apadula} \affiliation{\isu} \affiliation{\stonycrkp} 
\author{Y.~Aramaki} \affiliation{\riken} 
% ...
\collaboration{PHENIX Collaboration} \noaffiliation

\date{\today}


\begin{abstract}


PHENIX measurements are presented for the cross section and 
double-helicity asymmetry ($A_{LL}$) in inclusive $\pi^0$ production at 
midrapidity from $p$$+$$p$ collisions at $\sqrt{s}=510$~GeV from data 
taken in 2012 and 2013 at the Relativistic Heavy Ion Collider. The 
next-to-leading-order perturbative-quantum-chromodynamics theory 
calculation is in excellent agreement with the presented cross section 
results. The calculation utilized parton-to-pion fragmentation functions 
from the recent DSS14 global analysis, which prefer a smaller 
gluon-to-pion fragmentation function.  The $\pi^{0}A_{LL}$ results follow 
an increasingly positive asymmetry trend with $p_T$ and $\sqrt{s}$ with 
respect to the predictions and are in excellent agreement with the latest 
global analysis results.  This analysis incorporated earlier results on 
$\pi^0$ and jet $A_{LL}$, and suggested a positive contribution of gluon 
polarization to the spin of the proton $\Delta G$ for the gluon momentum 
fraction range $x>0.05$. The data presented here extend to a currently 
unexplored region, down to $x\sim0.01$, and thus provide additional 
constraints on the value of $\Delta G$.  

\end{abstract}

	
\maketitle


In the late 1980s, the EMC experiment~\cite{Ashman:1989ig} showed that the 
spins of quarks and anti-quarks might contribute only a fraction of the 
proton spin (about 1/3 from the recent global analyses of world spin 
polarized scattering 
data~\cite{deFlorian:2008mr,deFlorian:2009vb,Blumlein:2010rn,Leader:2010rb,Ball:2013lla}). 
This sparked several decades of world-wide effort to understand the proton 
spin structure in terms of quark and gluon polarizations and their orbital 
angular momentum, as evidenced by experimental programs at CERN, SLAC, 
DESY, JLAB, and BNL.

A key component of the Relativistic Heavy Ion Collider (RHIC) Spin program 
is the determination of the gluon spin contribution to the spin of the 
proton. High energy polarized proton collisions provide direct access to 
the gluon polarization $\Delta G$ within the proton through several gluon 
dominated hard scattering processes, such as high $p_T$ jet and hadron 
production~\cite{Bunce:2000uv}.  RHIC results on the double helicity 
asymmetry $A_{LL}$ in inclusive $\pi^0$ production at $\sqrt{s}=62.4$ and 
200~GeV from 
PHENIX~\cite{Adler:2006bd,Adare:2007dg,Adare:2008qb,Adare:2008aa} and jet 
production at $\sqrt{s}=200$~GeV from 
STAR~\cite{Abelev:2006uq,Abelev:2007vt} have made a significant 
contribution to the $\Delta G$ determination 
\cite{deFlorian:2008mr,deFlorian:2009vb}. Inclusion of the recent RHIC 
results from $\sqrt{s}=200$ GeV data collected in 2009 
\cite{Adare:2014hsq,Adamczyk:2014ozi} in the global next-to-leading-order 
(NLO) perturbative-quantum-chromodynamics (pQCD) analysis provided 
evidence for positive gluon polarization within the proton, with the 
integral of $\Delta G(x,Q^2=10$~GeV$^2)$ in the gluon momentum fraction 
$x>0.05$ being $0.20^{+0.06}_{-0.07}$ at 90\% 
C.L.~\cite{deFlorian:2014yva}. The RHIC high luminosity data at 
$\sqrt{s}=510$ GeV allow probing $\Delta G$ in the overlapping $x$ range 
at higher momentum transfer, and extends our understanding of $\Delta G$ 
to the unexplored lower $x$ region.

In this Letter, we present the PHENIX $\pi^0$ $A_{LL}$ results at 
$\sqrt{s}=510$ GeV from the RHIC 2012 and 2013 data sets, with an 
integrated luminosity of 20 and 108 pb$^{-1}$, respectively. We also 
present and discuss our results on $\pi^0$ unpolarized cross section 
measurements, which serve as an important test for the applicability of 
the NLO pQCD theory calculations in the accessed kinematic range. The 
theory is used to connect the measured asymmetries to gluon polarization 
in the proton~\cite{deFlorian:2008mr,deFlorian:2009vb,deFlorian:2014yva}.

The PHENIX experimental setup is described elsewhere~\cite{Adcox:2003zm}. 
In this analysis, $\pi^0$s were reconstructed via $\pi^0 \rightarrow 
\gamma\gamma$ decays using a highly-segmented electromagnetic calorimeter 
(EMCal), covering a pseudorapidity range of $|\eta| < 0.35$. The EMCal 
comprises two calorimeter types, a lead-scintillator (PbSc) sampling 
calorimeter and a lead-glass (PbGl) \v{C}erenkov calorimeter, with 
granularity $\Delta \eta \times \Delta \phi \sim 0.011 \times 0.011$ and 
$0.008 \times 0.008$, respectively.  Eight EMCal sectors (six PbSc and two 
PbGl) are located in two nearly back-to-back arms each covering $\Delta 
\phi \sim 90^{\circ}$ in azimuth. The PHENIX EMCal also generates a high 
$p_T$ photon (HPP) trigger when the deposited energy in any set of $4 
\times 4$ towers exceeds a pre-defined threshold.  Thin multiwire 
proportional chambers located in front of the EMCal were used as a veto to 
suppress the charged hadron background in $\pi^0$ reconstruction 
\cite{Adare:2014hsq}.  Beam-beam counters (BBC), positioned at $\pm$144 cm 
from the nominal interaction point along the beam line and covering 
$\eta=\pm$\,3.0--3.9, defined the minimum-bias (MB) collision trigger 
and determined the location of the collision vertex. Only events with 
collision vertices within $\pm$10 cm ($\pm$30 cm) of the nominal 
interaction point were used in the cross section (asymmetry) analysis. The 
BBCs were also used to calculate the integrated luminosity of the 
collected data sample and relative luminosity between colliding bunches 
with different spin configurations.  Zero-degree calorimeters (ZDC), 
located at $\pm18$ m and covering $|\eta|>6$, were used as another 
relative luminosity monitor. Equipped with a shower-maximum detector, the 
ZDC also provided monitoring of the transverse polarization component of 
colliding bunches in the PHENIX interaction region, utilizing the 
azimuthal asymmetry in forward neutron production in transversely 
polarized $p$$+$$p$ collisions~\cite{Adare:2012vw}.

\begin{figure}[!htb]
\includegraphics[width=0.8\linewidth]{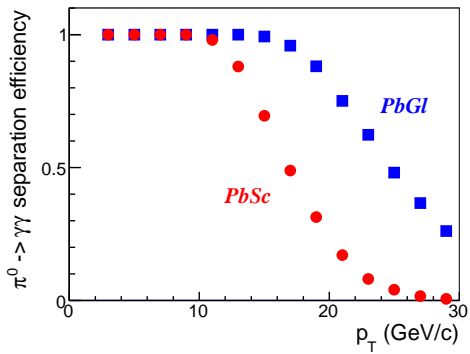}
\caption{\label{fig:merge} (color online)
The probability for two photons from $\pi^0$ decay to be separated by the 
PHENIX EMCal clustering algorithm vs $\pi^0$ $p_T$; 
obtained from {\sc geant}~\cite{geant} simulation for the two-photon energy 
asymmetry cut $\alpha<0.8$.
}
\end{figure}

As described in detail in Ref.~\cite{Adare:2007dg}, $\pi^0$s were 
reconstructed from two-photon invariant mass distributions. A time of 
flight cut and shower profile evaluation (energy distribution among EMCal 
towers) were used for photon identification. A minimal photon energy cut 
of 0.3 GeV and an energy asymmetry between the two photons 
$\alpha=|E1-E2|/(E1+E2)<0.8$ were applied. The $\pi^0$ peak width in the 
invariant mass distribution varied between 9 and 12 MeV/$c^2$ over the 
measured $p_T$ range. The resulting background fraction in the mass window 
of $\pm 25$ MeV/$c^2$ around the $\pi^0$ peak varied from $\sim$20\% at 
$p_T\sim$2~GeV/$c$ to $<$8\% at $p_T>5$~GeV/$c$. The two decay photons 
start merging in the PbSc (PbGl) EMCal at $\pi^0$ $p_T>10$ GeV/$c$ ($>15$ 
GeV/$c$). A 50\% merging probability is reached at $p_T\sim17$~GeV/$c$ 
(25~GeV/$c$) in the PbSc (PbGl), as shown in Fig.~\ref{fig:merge}.  For 
$p_T>24$~GeV/$c$, the majority of photon pairs are merged in the PbSc; in 
this $p_T$ range, only the PbGl data were used.

The invariant differential cross section for $\pi^0$ production is
calculated as
\begin{equation}
E \frac{d^3 \sigma}{dp^3} = \frac{1}{\cal L}
                            \cdot
                            \frac{1}{2 \pi p_T^{*}} 
                            \cdot
                            \frac{C \cdot N}{\Delta p_T \Delta y},
\label{eq:cross}
\end{equation}
where $N$ is the number of $\pi^0$'s observed in a $\Delta p_T$ wide bin at 
$p_T^{*}$ defined as the $p_T$ for which the cross section equals its average 
over the bin; $\Delta y$ is the rapidity range; $C$ includes corrections for 
trigger efficiency, geometrical acceptance, $\pi^0$ reconstruction 
efficiency, and detector resolution effects; $\cal L$ is the integrated 
luminosity for the analyzed data sample.

Two data samples were used for the $\pi^0$ cross section measurements, one 
collected with a MB trigger and the other with the HPP in coincidence with MB 
trigger. The MB trigger efficiency was obtained from the data collected with 
a dedicated HPP trigger operated without coincidence with MB trigger, and 
found to be $0.91\pm0.01$ independent of $p_T$. It accounts for the fact that 
only a fraction of inelastic $p$$+$$p$ collisions producing $\pi^0$ meson(s) 
fires the MB trigger. The HPP trigger efficiency vs $p_T$ was calculated in 
each arm separately from a set of events triggered by a high energy cluster 
in the opposite arm. It showed a characteristic threshold behavior with 
efficiency increasing from $\sim$1\% at $p_T=2$ GeV/$c$ to 93\% at $p_T>8$ 
GeV/$c$. For the cross section calculation, the MB triggered data sample was 
used at $p_T<6$ GeV/$c$, and HPP triggered data sample at higher $p_T$.

The reconstructed $\pi^0$ yields in each $p_T$ bin were corrected 
for geometrical acceptance, reconstruction efficiencies 
(e.g. due to the two-photon energy asymmetry cut), and smearing effects 
(due to the finite detector resolutions). 
The corrections were calculated with a simulation
containing the EMCal geometry, known detector inefficiencies, and photon 
energy and position smearing based on the known EMCal resolutions. 

The major systematic uncertainties in the $\pi^0$ cross section measurement 
are the energy scale (1.2\% uncertainty in the EMCal energy calibration 
translates to $\sim$7\% in cross section uncertainty), 
energy nonlinearity (up to 10\% for cross section depending on $p_T$), 
and merging corrections (up to 30\% in the bins with the highest 
probability for two photons to merge).
The large uncertainty at high $p_T$ reflects the sensitivity 
of the merging correction to shower-shape fluctuations and 
background conditions for asymmetric two-photon decays, having 
higher probability to survive the merging in the EMCal.
The other uncertainties, contributing $<$6\% altogether, 
are related to $\pi^0$ yield extraction 
and background subtraction, trigger efficiencies, 
geometrical acceptance calculation, smearing corrections, 
and photon conversion.
The uncertainties are assigned separately for the PbSc and the PbGl 
measurements.

A comparison of the results obtained from the PbSc and the PbGl 
is a key cross check, 
because the two calorimeters have a different response to hadrons 
(hence different background contamination in $\pi^0$ reconstruction),
and considerably different merging corrections versus $p_T$. 
The $\pi^0$ cross section results from the PbSc and the PbGl  
were in agreement within uncertainties in the overlapping $p_T$ range. 
The final spectrum was obtained from the combined PbSc and PbGl results, 
while for $p_T>24$ GeV/$c$ the PbGl results were used.
The total systematic uncertainties associated with the results 
vary from 8--10\% 
at $p_T<14$ GeV/$c$ to $\sim$30\% at the highest $p_T$.

The integrated luminosity ${\cal L}$ in Eq.(\ref{eq:cross}) was calculated 
from the accumulated number of MB triggers in the analyzed data sample 
normalized by the cross section of the processes firing the MB trigger 
in $p$$+$$p$ collisions. 
Similar to our previous analyses~\cite{Adare:2008qb,Adler:2003pb}, 
the cross section was defined using 
a vernier scan technique and found to be 32.5 mb with $\pm$10\% uncertainty.

In the 2013 RHIC run, the instantaneous luminosity delivered to PHENIX was 
so high that up to a third of all bunch crossings had more than one 
$p$$+$$p$ collision. To correct for this multiple-collision effect, 
we studied the ratio of the $\pi^0$ yield to the number of MB triggers 
(which is proportional to the measured $N/\cal L$ in Eq.(\ref{eq:cross}))
as a function of instantaneous MB trigger rate.

\begin{figure}[!htb]
\includegraphics[width=1.0\linewidth]{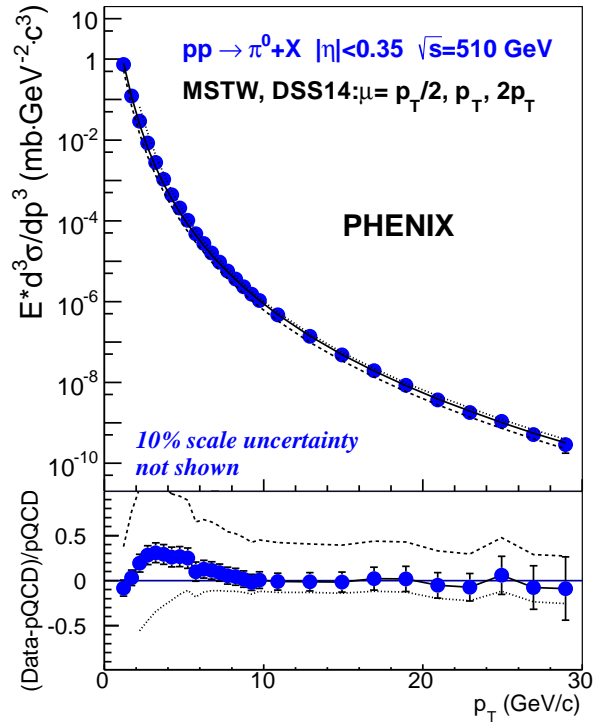}
\caption{\label{fig:cross} (color online) 
The neutral pion production cross section at midrapidity in $p$$+$$p$ 
collisions at $\sqrt{s}=510$ GeV as a function of $p_{T}$ and NLO pQCD 
calculations for theory scales $\mu=p_T/2$ (dotted line), $p_T$ (solid 
line) and $2p_T$ (dashed line), with $\mu$ representing equal 
factorization, renormalization, and fragmentation scales.  Note that the 
error bars, representing the combined statistical and point-to-point 
systematic uncertainties, are smaller than the points.  (bottom panel)  
Relative difference between the data and theory for the three theory 
scales. Experimental uncertainties are shown for the $\mu=p_T$ curve.
}
\end{figure}

Figure~\ref{fig:cross} shows the $\pi^0$ cross section versus $p_T$ 
compared to NLO pQCD calculations performed with MSTW~\cite{Martin:2009iq} 
parton distribution functions (PDF) and DSS14~\cite{deFlorian:2014xna} 
fragmentation functions (FF).  Compared to earlier FF analysis 
\cite{deFlorian:2007hc} the DSS14 recent global fit results preferred a 
smaller fraction of pions produced from gluon hadronization, driven mainly 
by the latest data from the Large Hadron Collider.  This theoretical 
calculation is in excellent agreement with the presented data.


In 2012 and 2013, RHIC provided PHENIX with colliding bunches of 
longitudinally polarized protons at $\sqrt{s}=510$ GeV. The bunch spin 
pattern was predefined in such a way that the colliding bunch pair 
helicity state alternated every bunch crossing, spaced 106 ns apart. This 
greatly suppressed the possibility of false asymmetries between colliding 
bunches with different helicity configuration, due to variation in 
detector performance. To remove possible systematic effects associated 
with particular bunch(es) in the process of filling, ramping up and 
storing the beams in RHIC rings, eight bunch spin patterns were used 
alternating every RHIC store, typically lasting eight hours. Beam 
polarizations were measured by RHIC polarimeters~\cite{cnipol} 
three-to-four times during the store.  For the two RHIC collider rings, 
labeled ``Blue'' (B) and ``Yellow'' (Y), the luminosity-weighted average 
polarizations in 2012 (2013) were $\langle P_B \rangle=0.55\pm0.02$ 
($0.55\pm0.02$) and $\langle P_Y \rangle=0.57\pm0.02$ ($0.56\pm0.02$). The 
degree of longitudinal polarization in the PHENIX interaction region 
was monitored by local polarimeters, based on the ZDC and shower-maximum 
detectors, which measured the residual transverse polarization of 
colliding bunches. The longitudinal component $P_L/P$ in both 2012 and 
2013 was $>0.998$, for both RHIC rings.

The $\pi^0$ $A_{LL}$ analysis technique is described in detail in 
Ref.~\cite{Adare:2014hsq}. The $A_{LL}$ for inclusive $\pi^0$ production, 
defined as the difference between cross sections for colliding bunches 
with the same helicity and opposite helicity, divided by the sum, is 
experimentally calculated as
\begin{equation}
A_{LL}^{\pi^0} = \frac{1}{P_B \cdot P_Y} \cdot \frac{N_{++}-R\cdot
N_{+-}}{N_{++} + R \cdot N_{+-}};~~~R=\frac{L_{++}}{L_{+-}},
\label{eq:a_ll}
\end{equation}
where $N$ is the number of $\pi^0$'s from the colliding bunches 
with the same ($++$) and opposite ($+-$) helicities, $R$ is the relative 
luminosity between bunches with the same and opposite helicities, and
$P_B$ and $P_Y$ are the two RHIC beam polarizations.

The $\pi^0$ yields were extracted from the HPP triggered sample in which 
the maximal energy photon of each pair candidate was explicitly required 
to fire the HPP trigger.  This test, along with a time-of-flight cut, 
suppressed the possibility of contamination from the neighboring bunch 
crossings to a negligible level. As in the cross section analysis, the 
$\pi^0$ candidates were counted within a $\pm 25$ MeV/$c^2$ window around 
the $\pi^0$ peak in the two-photon invariant mass distribution. The 
$A_{LL}$ was then corrected for the background $A_{LL}$ measured in the 
side bands on either side of the $\pi^0$ peak; this background asymmetry 
was found to be consistent with zero in all $p_T$ bins.

The relative luminosity $R$ was defined from the number of MB triggers in 
each bunch crossing, and cross checked using the number of collisions 
firing the ZDCs on both sides of the IR. The pile-up correction due to the 
high collision rate had a negligible effect on $R$. The resulting 
contribution of the relative luminosity uncertainty to $A_{LL}^{\pi^0}$ 
for the 2012 (2013) data was $\delta A_{LL}^{\pi^0}|_R=2.0\times10^{-4}$ 
($3.8\times10^{-4}$), affecting all $p_T$ bins in the same way.

$A_{LL}$ was measured for each PHENIX data taking segment (up to 90 
minutes long) to minimize the systematic effects from variation in $R$, 
beam polarization (decreasing during a store by $\Delta P$=0.005--0.010 
per hour), and HPP trigger performance. These asymmetries were averaged 
separately for the 2012 and 2013 data. Results from 2012 and 2013 were 
consistent within statistical uncertainties and the final result presented 
in this Letter is the average of these data sets.

The resulting $\pi^0$ $A_{LL}$ systematic uncertainties are (a) a 
correlated uncertainty from relative luminosity of $3.6\times10^{-4}$, (b) 
a correlated uncertainty from polarization measurements of 6.5\% (scale 
uncertainty), and (c) point-to-point uncertainty from background fraction 
determination under the $\pi^0$ peak in the two-photon invariant-mass 
distribution. The point-to-point uncertainties were found to be smaller 
than 10\% of the statistical uncertainty in all $p_T$ bins. As in the 
previous PHENIX analysis~\cite{Adare:2014hsq}, the contribution of other 
potential sources of systematic uncertainties was negligible.

\begin{figure}[!htb]
\includegraphics[width=1.0\linewidth]{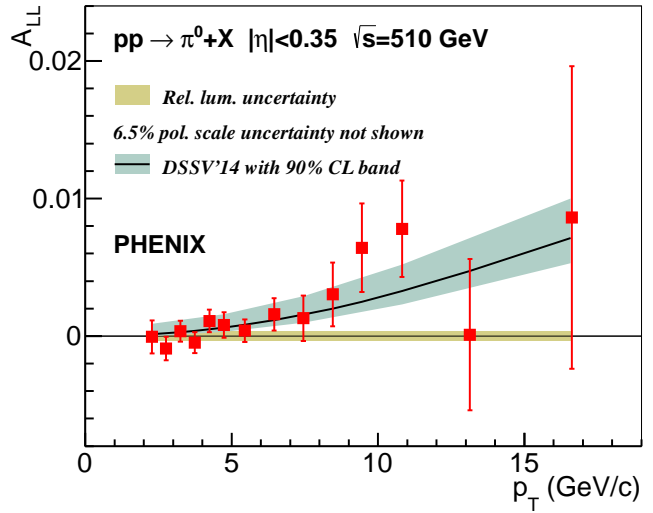}
\caption{\label{fig:all} (color online) 
$A_{LL}$ vs $p_T$ for $\pi^0$ production at midrapidity in $p$$+$$p$ 
collisions at $\sqrt{s}=510$ GeV.  Error bars are combined statistical and 
point-to-point systematic uncertainties.  The $A_{LL}=0$ (yellow) line is 
uncertainty from relative luminosity.  The theoretical curve with 90\% 
C.L. band (green) is from a DSSV14 
calculation~\protect\cite{deFlorian:2014yva}.
}
\end{figure}
\begin{figure}[!htb]
\includegraphics[width=1.0\linewidth]{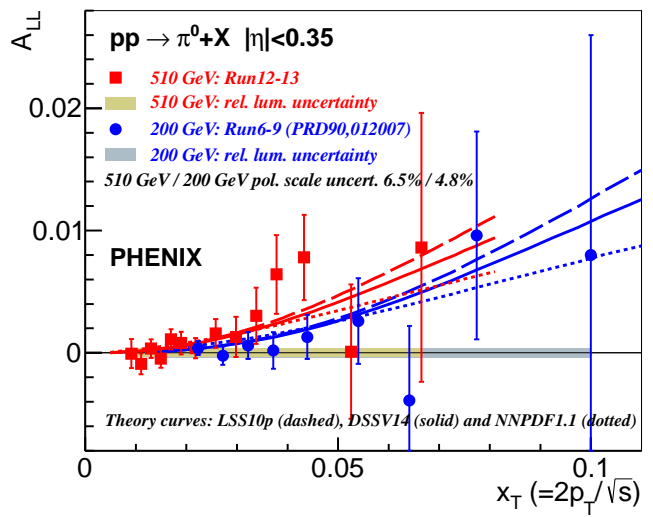}
\caption{\label{fig:alls} (color online)
$A_{LL}$ vs $x_T$ for $\pi^0$ production at midrapidity at 
$\sqrt{s}=200$~GeV (blue) from~\cite{Adare:2014hsq} and 510 GeV (red) from 
this analysis.  Error bars are combined statistical and point-to-point 
systematic uncertainties. Note that the relative luminosity uncertainties 
from two data samples are about the same, hence are indistinguishable in 
the plot in the overlapping $x_T$ range.  Theoretical curves are from 
recent NLO global 
analyses~\protect\cite{deFlorian:2014yva,Leader:2010rb,Ball:2013lla}, with 
the lower curves (blue) for $\sqrt{s}=200$~GeV and the higher curves (red) 
for $\sqrt{s}=510$~GeV.
}
\end{figure}

Figure~\ref{fig:all} shows the $\pi^0$ $A_{LL}$ asymmetries at 
$\sqrt{s}=510$ GeV compared with the DSSV14 calculation 
\cite{deFlorian:2014yva} based on a global fit of the world helicity 
asymmetry data. Comparing the data to the DSSV14 curve we obtain 
$\chi^2$/NDF=8.0/14, while comparing to the $A_{LL}=0$ hypothesis we 
obtain $\chi^2$/NDF = 18.2/14; the data prefer the DSSV14 curve by a 
little more than 3 standard deviations.

Figure~\ref{fig:alls} shows $\pi^0$ $A_{LL}$ data from PHENIX at both 
$\sqrt{s}=200$ GeV [14] and 510 GeV, along with NLO pQCD analyses from 
three groups~\cite{deFlorian:2014yva,Leader:2010rb,Ball:2013lla}. All 
three analyses predict an increase in $\pi^0$ $A_{LL}$ at the same $x_T$ 
due to pQCD evolution, with $x_T=2p_T/\sqrt{s}$. Our data is consistent 
with such an increase.

In summary, we have presented the unpolarized cross section and double 
helicity asymmetry for $\pi^0$ production at midrapidity for $p$$+$$p$ 
collisions at $\sqrt{s}=510$ GeV. The NLO pQCD calculation is in excellent 
agreement with the presented cross section results. The calculation 
utilized the recent DSS14 set of fragmentation functions, which prefer the 
reduced fraction of pions produced from gluon hadronization. The $\pi^0$ 
$A_{LL}$ results follow a positive asymmetry trend with $p_T$ and 
$\sqrt{s}$ predicted by NLO pQCD and are in excellent agreement with the 
latest global fit results, which suggested a nonzero gluon polarization in 
the proton for the gluon momentum fraction range $x>0.05$. These global 
fit results included RHIC $\pi^0$ $A_{LL}$ data at $\sqrt{s}=62.4$ GeV and 
200 GeV and jet $A_{LL}$ data at $\sqrt{s}=200$ GeV. 
The presented data at $\sqrt{s}=510$ GeV extend the $x$ range probed down 
to $x\sim0.01$ and provide an additional constraint on $\Delta G$ in this 
$x$ range~\cite{sassot}, which is a crucial step in the nearly two decades 
of world-wide efforts to understand the contribution of gluon polarization 
to the spin of the proton.  We note the recent $\pi^0$ $A_{LL}$ results at 
$\sqrt{s}=200$~GeV and forward pseudorapidity $0.8<\eta<2$ from STAR 
covering the gluon $x$ range down to $x\sim0.01$ (although with large 
uncertainties)~\cite{Adamczyk:2013yvv}.  Data collected by PHENIX with 
forward EMCal at pseudorapidity $3.1<\eta<3.9$ and $\sqrt{s}=510$~GeV will 
further extend the $x$ range probed down to $x\sim0.001$.



\begin{acknowledgments}

We thank the staff of the Collider-Accelerator and Physics
Departments at Brookhaven National Laboratory and the staff of
the other PHENIX participating institutions for their vital
contributions.  We also thank Marco Stratmann for providing
the theoretical calculations and for informative discussions.
We acknowledge support from the 
Office of Nuclear Physics in the
Office of Science of the Department of Energy,
the National Science Foundation, 
Abilene Christian University Research Council, 
Research Foundation of SUNY, and
Dean of the College of Arts and Sciences, Vanderbilt University 
(U.S.A),
Ministry of Education, Culture, Sports, Science, and Technology
and the Japan Society for the Promotion of Science (Japan),
Conselho Nacional de Desenvolvimento Cient\'{\i}fico e
Tecnol{\'o}gico and Funda\c c{\~a}o de Amparo {\`a} Pesquisa do
Estado de S{\~a}o Paulo (Brazil),
Natural Science Foundation of China (P.~R.~China),
Croatian Science Foundation and
Ministry of Science, Education, and Sports (Croatia),
Ministry of Education, Youth and Sports (Czech Republic),
Centre National de la Recherche Scientifique, Commissariat
{\`a} l'{\'E}nergie Atomique, and Institut National de Physique
Nucl{\'e}aire et de Physique des Particules (France),
Bundesministerium f\"ur Bildung und Forschung, Deutscher
Akademischer Austausch Dienst, and Alexander von Humboldt Stiftung (Germany),
National Science Fund, OTKA, K\'aroly R\'obert University College, 
and the Ch. Simonyi Fund (Hungary),
Department of Atomic Energy and Department of Science and Technology (India), 
Israel Science Foundation (Israel), 
Basic Science Research Program through NRF of the Ministry of Education (Korea),
Physics Department, Lahore University of Management Sciences (Pakistan),
Ministry of Education and Science, Russian Academy of Sciences,
Federal Agency of Atomic Energy (Russia),
VR and Wallenberg Foundation (Sweden), 
the U.S. Civilian Research and Development Foundation for the
Independent States of the Former Soviet Union, 
the Hungarian American Enterprise Scholarship Fund,
and the US-Israel Binational Science Foundation.

\end{acknowledgments}



%
 
\end{document}